# Quantum interference with slits

**Thomas V Marcella**

Department of Physics and Applied Physics, University of Massachusetts Lowell, Lowell, MA 01854-2881, USA

E-mail: thomasmarcella@verizon.net

In the experiments considered here, we measure the y-component of momentum for a particle passing through a system of slits. The source-slit system is the preparation apparatus that determines the state vector. Recognizing that a system of slits is a position-measuring device allows us to ascertain that the state vector is a position state. Then, writing the state vector in momentum space provides a straightforward calculation for the probability amplitude and its corresponding probability function. Interference effects, if any, are inherent in the probability function We determine the statistical distribution of scattered particles for four different slit systems. The results are in agreement with the well-known interference patterns obtained in classical wave optics.

# 1. Introduction

The double-slit experiment is the archetypical system used to demonstrate quantum mechanical behavior. It is said by Feynman [1] to "contain the only mystery" in all of quantum mechanics. Numerous textbooks and journal articles discuss slit interference, usually in conjunction with wave-particle duality.

Most authors emphasize that classical physics cannot describe the double slit experiment with particles. Yet, bolstered by the deBroglie hypothesis, they still ascribe to the classical maxim, "Waves exhibit interference. Particles do not." They then conclude that, "When particles exhibit interference, they are behaving like waves". Then the subsequent analysis is simply wave theory, and any interference effects are made to agree with Young's experiment. Thus, classical wave optics, rather than quantum mechanics, is used to explain quantum interference. For example, Ohanian [2] states " ----the maxima of this interference pattern are given by a formula familiar from wave optics."

Some authors do suggest that a quantum mechanical approach is lacking. Liboff [3] tells us, "The first thing to do is to solve Schroedinger's equation and calculate $|\psi|^2$ at the screen." Ballentine [4] makes a similar statement when discussing diffraction from a periodic array: "--------solve the Schroedinger equation with boundary conditions corresponding to an incident beam from a certain direction, and hence determine the position probability density $|\Psi(\vec{x})|^2$ at the detectors." But he then says, "An exact solution of this equation would be very difficult to obtain, --------". The difficulty according to Merzbacher [5] is that, "A careful analysis of the interference experiment would require detailed consideration of the boundary conditions at the slits." In spite of these misgivings, quantum mechanics does provide a straightforward calculation for the probability distribution of the scattered particles.

Quantum mechanics is a theory about observables and their measurement. Its postulates provide, among other things, a set of instructions for calculating the probability of obtaining a particular result when an observable is measured. These probability calculations require a state vector $|\psi\rangle$, which is determined by the preparation procedure. Its representation is dictated by the observable being measured;

$$|\psi\rangle = \sum |a_k\rangle\langle a_k|\psi\rangle. \tag{1}$$

The basis vectors $|a_k\rangle$ are the eigenvectors of the measured observable $\hat{A}$. Having obtained the state vector $|\psi\rangle$, the probability that a measurement of observable $\hat{A}$ yields the value $a_k$ is given by the Born postulate

$$P_k = |\langle a_k|\psi\rangle|^2. \tag{2}$$

The state vector $|\psi\rangle$ and the probability distribution $|\langle a_k|\psi\rangle|^2$ are unique for a given experiment. State preparation and measurement are discussed at length in Ballentine [4].

We expect, then, that a quantum mechanical description of a slit experiment will

a) clearly define which observable is being measured,

b) describe the preparation procedure that determines the state vector, and

c) yield the probability function for the scattered particles.

This author is unaware of any description of slit interference based on the formalism of quantum mechanics other than the work of Barut and Basri [6], who used the path integral method of Feynman.

## 2. Probability functions and quantum interference

The experiment considered here consists of the apparatus shown in figure 1. For such an experiment, the source-slit system, which determines the possible $y$-coordinate(s) of the particle at the slits, is the preparation apparatus. Thus, the state vector is a position state.

Because position and momentum are non-commuting observables, a particle passing through slits always has an uncertainty in its $y$-component of momentum. It can be scattered with any one of the continuum of momentum eigenvalues $p_y = p\sin\theta$, where $-\pi/2 \leq \theta \leq \pi/2$. Measurement of a well-defined scattering angle $\theta$ constitutes a measurement of the observable $\hat{p}_y$ and, therefore, the basis vectors in Hilbert space are the momentum eigenvectors $|p_y\rangle$.

The probability that a particle leaves the slit apparatus with momentum $p_y$ is, then,

$$P(p_y) = \left|\langle p_y | \psi \rangle\right|^2. \tag{3}$$

It is this probability function that exhibits quantum interference. Its maxima and minima, if any, correspond to constructive and destructive interference respectively.

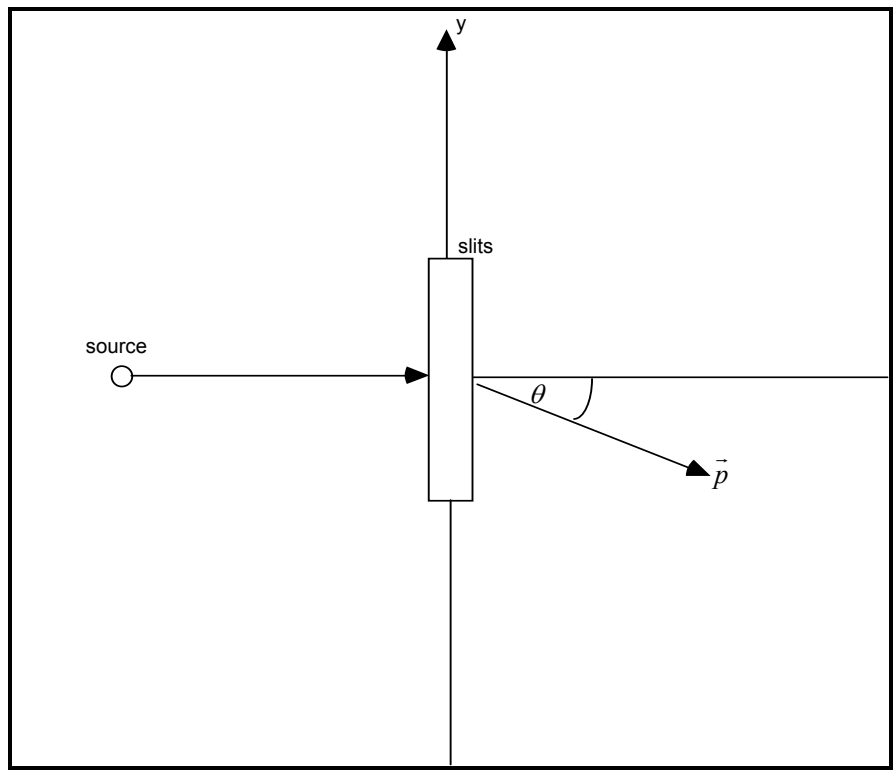

**Figure 1.**  Particle scattering from slits. A particle originating at the source is scattered at angle $\theta$ by the system of slits. A particle passing through the slits has undergone a unique state preparation that determines the probability of scattering at angle $\theta$.

In the position representation, the free-particle momentum eigenfunction corresponding to the eigenvalue $p_y$ is

$$\langle y | p_y \rangle = e^{i(p_y/\hbar)y} / \sqrt{2\pi}, \tag{4}$$

and the probability amplitude for scattering with momentum $p_y$ is

$$\langle p_y | \psi \rangle = \int_{-\infty}^{\infty} \langle p_y | y \rangle \langle y | \psi \rangle \, dy = \int_{-\infty}^{\infty} e^{-i(p_y/\hbar)y} \psi(y) \, dy / \sqrt{2\pi}. \tag{5}$$

An examination of the corresponding probability function $P(p_y) = |\langle p_y | \psi \rangle|^2$ will ascertain whether or not there is interference.

In what follows, we evaluate the integral of equation (5) by first constructing the position state function $\langle y | \psi \rangle = \psi(y)$. We do this for four source-slit systems, including the double slit.

### 2.1. Scattering from a narrow slit

A narrow slit of infinitesimal width is an ideal measuring device; it determines the position with infinite resolution [7]. A particle emerging from a slit at $y = y_1$ is in the position eigenstate $|y_1\rangle$. In the position representation, the eigenfunction of position is the Dirac delta function

$$\psi(y) = \langle y | y_1 \rangle = \delta(y - y_1). \tag{6}$$

and the probability amplitude for a particle emerging from the slit with momentum $p_y$ is,

$$\langle p_y | \psi \rangle = \int_{-\infty}^{\infty} e^{-i(p_y/\hbar)y} \delta(y - y_1) \, dy \Big/ \sqrt{2\pi} = e^{-i(p_y/\hbar)y_1} / \sqrt{2\pi}. \tag{7}$$

The corresponding probability function is

$$P(p_y) = |\langle p_y | \psi \rangle|^2 = \left| e^{-i(p_y/\hbar)y_1} / \sqrt{2\pi} \right|^2 = \text{constant}. \tag{8}$$

It is equally probable that the particle is scattered at any angle. There is no interference.

### 2.2. Scattering from a double narrow slit

We again assume that the slits are infinitesimally thin. For such a double slit apparatus, the observable $\hat{y}$ has two eigenvalues, $y_1$ and $y_2$. Assuming the source-slit geometry does not favor one slit is over the other, the state vector is the superposition of position eigenvectors

$$|\psi\rangle = (|y_1\rangle + |y_2\rangle)/\sqrt{2}, \tag{9}$$

and
$$\psi(y) = \langle y|\psi\rangle = (\delta(y-y_1) + \delta(y-y_2))/\sqrt{2} \quad . \tag{10}$$

Here, the amplitude for finding the particle with momentum $p_y$ is

$$\langle p_y|\psi\rangle = \frac{1}{\sqrt{2\pi}}\left(\int_{-\infty}^{\infty} e^{-i(p_y/\hbar)y}\,\delta(y-y_1)\,dy + \int_{-\infty}^{\infty} e^{-i(p_y/\hbar)y}\,\delta(y-y_2)\,dy\right)\!\Big/\sqrt{2}$$

$$= \left(e^{-i(p_y/\hbar)y_1} + e^{-i(p_y/\hbar)y_2}\right)\!\Big/2\sqrt{\pi}. \tag{11}$$

From which we get the probability function

$$P(p_y) = |\langle p_y|\psi\rangle|^2 = \left(2 + e^{i(p_y/\hbar)(y_1-y_2)} + e^{-i(p_y/\hbar)(y_1-y_2)}\right)\!\Big/4\pi$$

$$= (1 + \cos(p_y/\hbar)d)/2\pi, \tag{12}$$

where $d = y_1 - y_2$ is the distance between the slits. We see that this probability function does have relative maxima and minima and quantum interference does occur. Using $p_y = p\sin\theta$ we obtain the angular distribution of scattered particles

$$P(\theta) = [1 + \cos(pd\sin\theta/\hbar)]/2\pi. \tag{13}$$

A plot of which is shown in figure 2.

If we define $\phi = pd\sin\theta/\hbar$ and use the half-angle formula $1 + \cos\phi = 2\cos^2(\phi/2)$, the probability function takes the form

$$P(\phi) = \cos^2(\phi/2)/\pi. \tag{14}$$

This is the familiar intensity distribution for Fraunhofer diffraction as given in Guenther [8], among others.

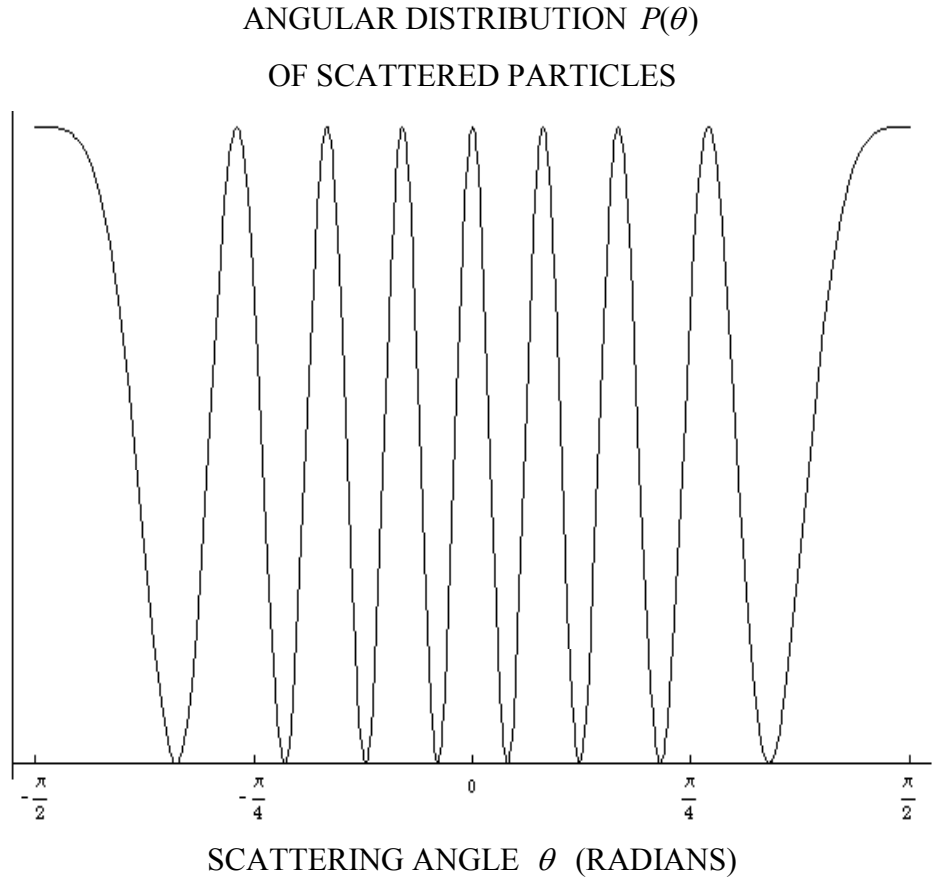

**Figure 2** Angular distribution of particles scattered from double narrow slits. The distance between the slits determines the number of interference fringes. In this example the distance between the slits is $d = 4\lambda$, where $\lambda$ is the deBroglie wavelength.

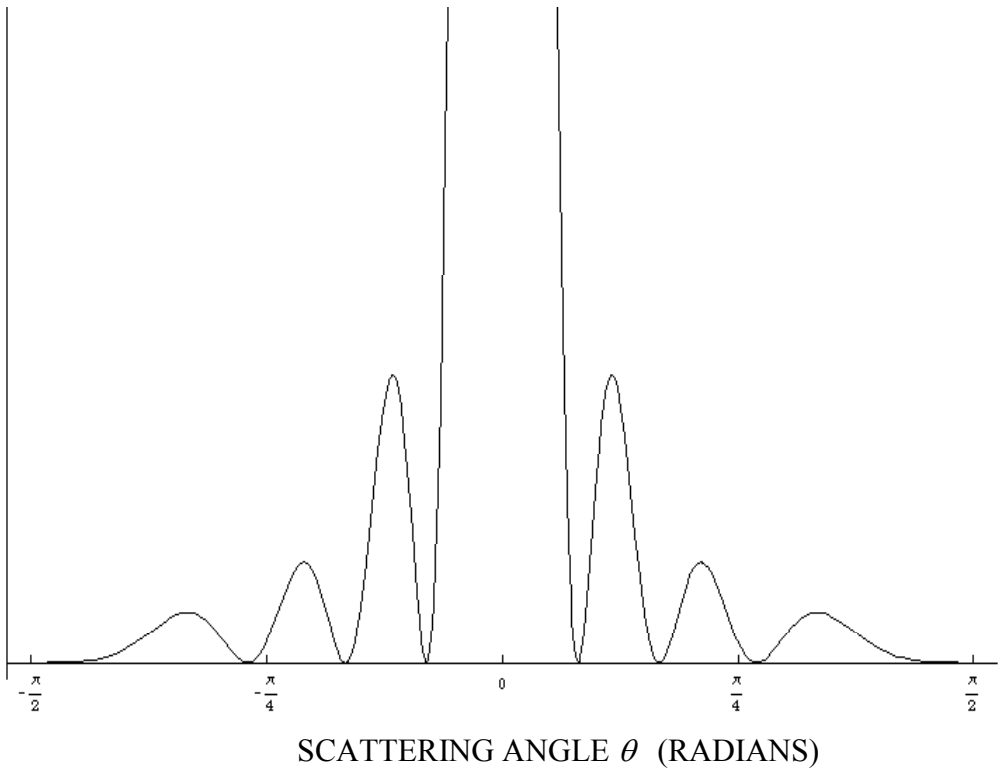

ANGULAR DISTRIBUTION $P(\theta)$
OF SCATTERED PARTICLES

SCATTERING ANGLE $\theta$ (RADIANS)

**Figure 3**  Angular distribution of particles scattered from a single slit of finite width. The number of interference fringes is determined by the slit width. In this example the slit width is $a = 4\lambda$, where $\lambda$ is the deBroglie wavelength.

**2.3. Scattering from a slit of finite width**

A slit of finite width is an imperfect apparatus for measuring position. It cannot distinguish between different position eigenvalues and a particle emerging from a slit of width $a$ can have any value of observable $\hat{y}$ in the continuum $-a/2 \leq y \leq a/2$. Assuming an equal probability of passing through the slit at any point, a particle at the slit is in the superposition state

$$\psi(y) = \langle y|\psi\rangle = \begin{cases} 1/\sqrt{a} & -a/2 \leq y \leq a/2 \\ 0 & \text{elsewhere} \end{cases}. \tag{15}$$

Here, the probability amplitude is

$$\langle p_y|\psi\rangle = \frac{1}{\sqrt{2\pi a}} \int_{-a/2}^{a/2} e^{-ip_y y/\hbar} dy = [i\hbar/p_y\sqrt{2\pi a}][e^{-iap_y/2\hbar} - e^{iap_y/2\hbar}]$$

$$= 2\hbar \sin(ap_y/2\hbar)/p_y\sqrt{2\pi a}, \tag{16}$$

and the corresponding probability function is

$$P(k_y) = |\langle k_y|\psi\rangle|^2 = [2/\pi a k_y^2]\sin^2(ak_y/2). \tag{17}$$

This result is shown in terms of the scattering angle $\theta$ in figure 3. We see that $P(p_y)$ is the well-known diffraction pattern for light if we define $\alpha = ap_y/2\hbar = ap\sin\theta/2\hbar$ and write

$$P(\alpha) = (a/2\pi)(\sin\alpha/\alpha)^2 \tag{18}$$

**2.4. Scattering from a double finite-width slit**

As a final example, we consider a particle passing through a double-slit apparatus consisting of two slits each of finite width $a$. This is a more realistic description of the double slit experiment. Here, the state vector is

$$|\psi\rangle = [|\psi_1\rangle + |\psi_2\rangle]/\sqrt{2}, \tag{19}$$

where

$$\langle y|\psi_1\rangle = \begin{cases} 1/\sqrt{a} & y_1 - a/2 \leq y \leq y_1 + a/2 \\ 0 & \text{elsewhere} \end{cases} \tag{20}$$

and

$$\langle y|\psi_2\rangle = \begin{cases} 1/\sqrt{a} & y_2 - a/2 \leq y \leq y_2 + a/2 \\ 0 & \text{elsewhere.} \end{cases} \tag{21}$$

Again, we calculate the probability amplitude

$$\langle p_y|\psi\rangle = [\langle p_y|\psi_1\rangle + \langle p_y|\psi_2\rangle]/\sqrt{2}$$

$$= \frac{1}{\sqrt{2\pi a}}\left[\int_{y_1-a/2}^{y_1+a/2} e^{-ip_y y/\hbar} dy + \int_{y_2-a/2}^{y_2+a/2} e^{-ip_y y/\hbar} dy\right]$$

$$= \left[e^{-ip_y(y_1+a/2)/\hbar} - e^{-ip_y(y_1-a/2)/\hbar} + e^{-ip_y(y_2+a/2)/\hbar} - e^{-ip_y(y_2-a/2)/\hbar}\right] i\hbar/p_y\sqrt{2\pi a}$$

$$\tag{22}$$

With a slight manipulation of terms this amplitude becomes

$$\langle p_y | \psi \rangle = 2\hbar \left[ \left( e^{-i p_y y_1 / \hbar} + e^{-i p_y y_2 / \hbar} \right) \sin(a p_y / 2\hbar) \right] / \left[ p_y \sqrt{2\pi a} \right] \quad (23)$$

and the corresponding probability distribution is

$$P(p_y) = 4\hbar^2 \left[ (1 + \cos(p_y d / \hbar)) \sin^2(a p_y / 2\hbar) \right] / \pi a p_y^2. \quad (24)$$

The angular distribution $P(\theta)$ is shown in figures 4 and 5 for two different slit configurations. Writing equation (23) in terms of $\phi = pd \sin\theta / \hbar$ and $\alpha = ap \sin\theta / 2\hbar$, we again get the optical form.

$$P(\phi) = 2a \left[ \cos^2(\phi/2)(\sin\alpha/\alpha)^2 \right] / \pi. \quad (25)$$

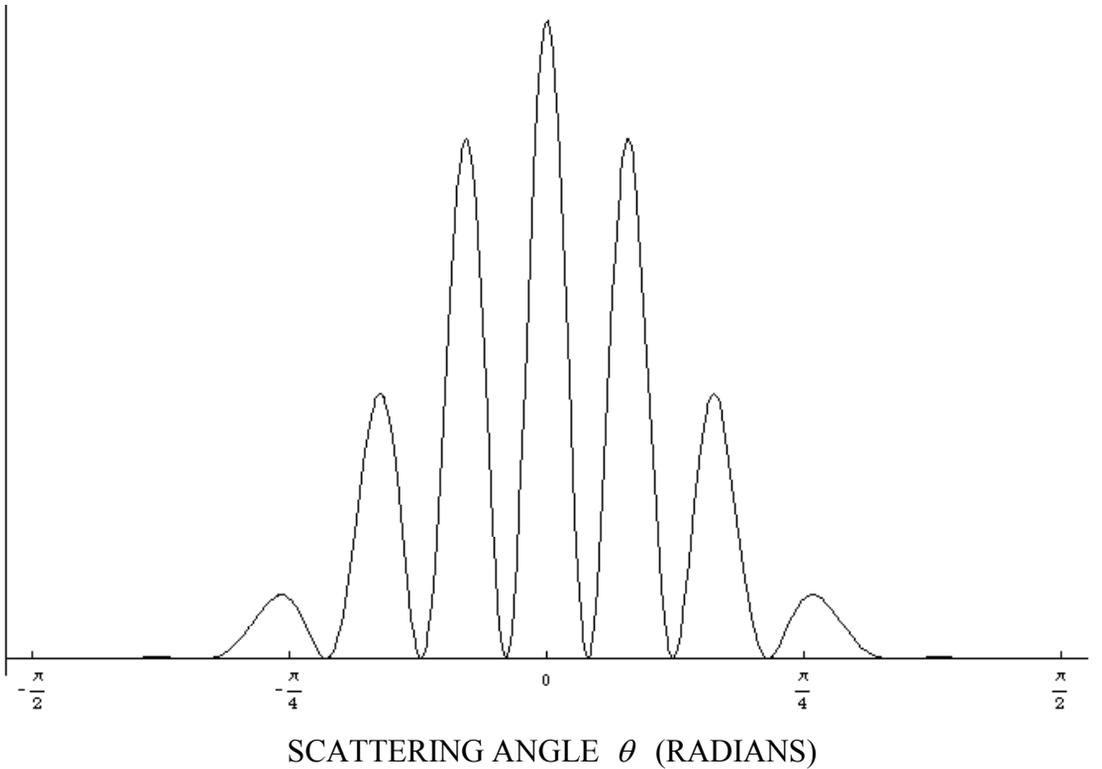

ANGULAR DISTRIBUTION $P(\theta)$
OF SCATTERED PARTICLES

SCATTERING ANGLE $\theta$ (RADIANS)

**Figure 4**   Angular distribution of particles scattered from two slits of finite width. Here, the slit width is $a = \lambda$ and the distance between the slits is $d = 4\lambda$, where $\lambda$ is the deBroglie wavelength.

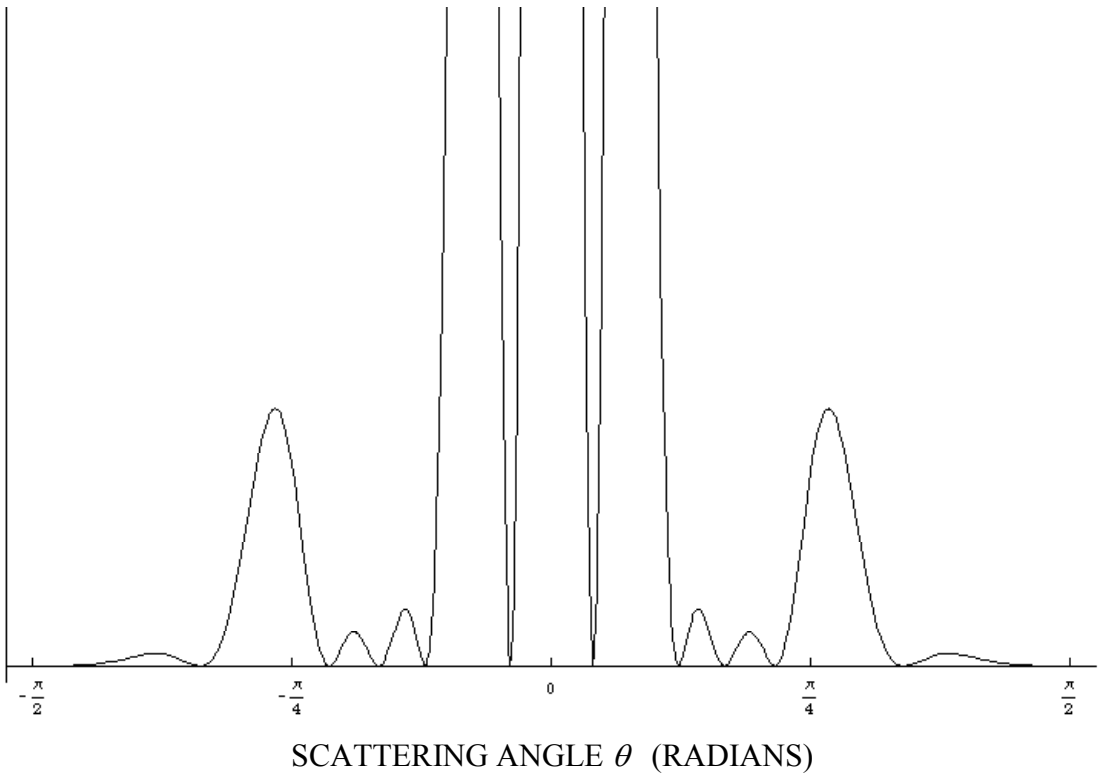

ANGULAR DISTRIBUTION $P(\theta)$

OF SCATTERED PARTICLES

SCATTERING ANGLE $\theta$ (RADIANS)

**Figure 5** Angular distribution of particles scattered from two slits of finite width. In this example, the slit width is $a = 2\lambda$ and the distance between the slits is $d = 4\lambda$, where $\lambda$ is the deBroglie wavelength

## 3. Concluding remarks

In this presentation, the Born postulate is used to obtain the interference pattern for particles scattered from a system of slits without referring, a priori, to classical wave theory. Having identified the state vector as a position state and the measured observable as the momentum, we obtain explicit expressions for the state vector $|\psi\rangle$ and its corresponding probability function $P(p_y) = |\langle p_y|\psi\rangle|^2$. The results are in agreement with wave optics.

Quantum interference can occur only when a large number of identically prepared particles are observed. These particles are detected at different locations, one at a time [9]. A single particle is always detected as a localized entity and no wave properties can be discerned from it.

It is interesting that for particles scattered from a double slit, the probability amplitude that gives rise to the interference is due to a superposition of delta functions.